\documentclass{aa}
\usepackage{color}
\usepackage{graphicx}
\usepackage{subcaption} 
\usepackage{tablefootnote}

\begin{document}

\title{Pulsar based modeling of point spread function of Fermi Large Area Telescope}
\author{Jeffrey Blunier$^1$, Andrii Neronov$^{1,2}$, Dmitri Semikoz$^1$}
\institute{Universit\'e Paris Cit\'e, CNRS, Astroparticule et Cosmologie, F-75013 Paris, France
\and
Laboratory of Astrophysics, \'Ecole Polytechnique F\'ed\'erale de Lausanne, CH-1015 Lausanne, Switzerland}

\authorrunning{Blunier, Neronov \& Semikoz}
\titlerunning{Pulsar based modeling of Fermi/LAT PSF}

\abstract{Sensitivity of searches for extended emission around gamma-ray sources is naturally limited by the precision of the knowledge of the Point Spread Function (PSF) of gamma-ray telescopes. Inaccuracies in the PSF models of the Fermi Large Area Telescope (LAT) can potentially lead to false positive detections of source extension.}{We explore uncertainties in the Fermi/LAT PSF by comparing the PSF models provided by the Fermi/LAT Instrument Response Functions (IRFs) with signals of bright pulsars.}{We compare the analytical PSF models of Fermi/LAT IRFs with pulsar data and fit the pulsar data with the same analytical model as in the Fermi/LAT IRFs to derive an improved set of PSF parameters. We then apply this revised PSF parameterisation to the search of extended emission around a blazar, Mrk 501.}{We find that the parameters of the analytical PSF models of Fermi/LAT IRFs are inconsistent with the pulsar data. We obtain an improved set of PSF parameters from the fits to pulsar data that is consistent with observations. We find no evidence of the previously reported extended signal around Mrk 501 if the revised PSF consistent with pulsar data is used in data analysis.}{}

\maketitle

\section{Introduction}
\label{sec:intro}

Extragalactic sources of TeV $\gamma$-rays may be surrounded by lower energy extended emission originating from electrons and positrons deposited in the intergalactic medium between the source and the Earth by interactions of the highest energy $\gamma$-rays with low-energy photons of the extragalactic background light \citep{Aharonian:1993vz,Neronov:2006lki}. The extended emission appears as an excess in the radial profiles of the source signal \citep{2009PhRvD..80l3012N,2013A&A...554A..31N} that is superimposed on the parent point source signal distributed according to the telescope Point Spread Function (PSF). Searches for such extended emission have so far mostly resulted in the upper limits on the extended emission flux. These upper limits have been used to derive a lower bound on the strength of Intergalactic Magnetic field (IGMF) \citep{2010Sci...328...73N,Blunier:2025ddu,MAGIC:2022piy}. A notable exception is the result of \cite{Webar:2025qbp} that has reported a detection of the extended emission around the second-brightest extragalactic TeV source on the sky, Mrk 501, a blazar type Active Galactic Nucleus (AGN). This extended emission is found in the data of Fermi Large Area Telescope (LAT). According to \cite{Webar:2025qbp}, such a detection corresponds to a specific combination of IGMF parameters: strength $B\sim 10^{-15}$~G and correlation length $\lambda_B\sim 10$~kpc.

Sensitivity of searches for the extended emission with both space-based Fermi/LAT and ground-based Imaging Atmospheric Cherenkov Telescopes \citep{HEGRA:2000ihg,2010Sci...328...73N,2010A&A...524A..77A,2014A&A...562A.145H,VERITAS:2017gkr,Webar:2025qbp} is limited by the precision of knowledge of the telescope PSF. Imperfections in the early modeling of the Fermi/LAT PSF have previously led to false positive detection(s) of extended emission around blazars \citep{Ando:2010rb}. At the initial stages of operation of Fermi/LAT telescope, the PSF modeling was based on Monte-Carlo simulations of $\gamma$-ray events that did not properly reproduce the observed event distributions. \cite{2011A&A...526A..90N} has reported a phenomenological study of Fermi/LAT PSF based on real data, that has revealed limitations of the Monte-Carlo modeling of the PSF. This study has concluded that the detection of extended emission claimed in \cite{Ando:2010rb} was spurious. The Fermi/LAT collaboration has performed an independent analysis of the PSF based on real telescope data \citep{2013ApJ...765...54A} and reached a similar conclusion. 
Most recently, the Fermi/LAT collaboration has re-derived the properties of the PSF from improved Monte-Carlo modeling\footnote{https://fermi.gsfc.nasa.gov/ssc/data/analysis/documentation /Cicerone/Cicerone\_LAT\_IRFs/IRF\_PSF.html} so that the Instrument Response Functions (IRF) released by the Fermi/LAT collaboration currently do not rely on the real data. The Fermi/LAT performance web-page\footnote{https://fermi.gsfc.nasa.gov/ssc/data/analysis/LAT\_caveats.html} specifies that a validation study of the PSF parameterisations has been performed and involved a comparison of the PSF models derived from the IRF parameters with the real data of Vela pulsar and of a set of AGN. This web page quotes a 5\% level uncertainty on the 68\% containment radius of the PSF in the 0.1 GeV to 10 GeV energy range, but no other tests. Useful as it is (see e.g. \cite{2013ApJ...765...54A}), the information on the 68\% signal containment radius is not sufficient for assessment of the PSF uncertainties in the context of the search of the extended emission around extragalactic sources and discovery of IGMF. It is possible to find two model PSF profiles that have the same 68\% containment radii but are largely different in shape (e.g. a Gaussian and an exponential function, or a profile with a power-law tail at large angular distances). 

In what follows we compare the Fermi/LAT PSF given by the IRF with observational data with the goal to assess the precision of the PSF parameters knowledge. In section \ref{sec:psf} we summarize the approach for description of the PSF in the Fermi/LAT IRF. In section \ref{sec:pulsars} we present our method of measurement of PSF uncertainties using the signals of bright pulsars. We show that the PSF models with parameters given in the IRF files do not properly describe the data. In Section \ref{sec:psf_revision} we present an improved PSF that is consistent with pulsar data. We use the improved PSF in re-analysis of Mrk 501 data in section \ref{sec:mrk}. Finally in section \ref{sec:discussion} we discuss implications of our results.

\section{Fermi/LAT PSF parameterisation}
\label{sec:psf}

Fermi-LAT IRFs use an analytical model of the PSF to describe the radial distributions of Monte-Carlo simulated events around sources. The PSF is modeled as a sum of two King functions,
\begin{equation}
K(x,\sigma,\gamma)=\frac{1}{2\pi\sigma^2}\left(1-\frac{1}{\gamma}\right)\left[1+\frac{1}{2\gamma}\frac{x^2}{\sigma^2}\right]^{-\gamma}
\label{eq:king}
\end{equation}
with different parameters $\sigma_c,\gamma_c$ for the "core" and $\sigma_t,\gamma_t$ for the "tail" components of the PSF:
\begin{equation}
PSF(x)=f_{c}K(x,\sigma_{c},\gamma_{c})+(1-f_{c})K(x,\sigma_{t},\gamma_{t})
\label{eq:psf}
\end{equation}
The factor $f_c$ describes the relative importance of the core and tail components in the overall PSF.

The parameters $\sigma_c,\gamma_c,f_c,\sigma_t,\gamma_t$ depend on the energy $E$ and off-axis angle $\theta$ of gamma-ray events detected by LAT. These parameters are stored as lookup tables (on a grid of $E$ and $\cos\theta$ values) in the IRF files for different event selections. The parameter $x=\psi/S_P(E)$ in (\ref{eq:psf}) is a re-scaled angular distance $\psi$ from the source with the scaling factor $S_P(E)$ that depends on energy (and on the event selection) 
\begin{equation}
S_P(E)=\sqrt{c_0^2\left(\frac{E}{100 \mbox{ MeV}}\right)^{-2\beta}+c_1^2}
\end{equation}
The re-scaling parameters $c_0, c_1,\beta$ are also stored in the IRF files. The PSF parameters for different $\gamma$-ray event types are stored in different IRF files. Overall, the PSF parameterisations are presented as lookup tables for a set of 23 logarithmically spaced energy bins between 10~MeV and 3~TeV, 8 linearly spaced bins of cosine of the off-axis angle between 0.2 and 1. The $\gamma$-ray events are divided into groups based on either type of interaction in the LAT detector (FRONT / BACK) or on the quality of reconstruction of the arrival direction (PSF0-PSF3). For the PSF0-PSF3 grouping, there is the total of 805 parameters, if one considers also the 69 parameters of the function $S_P(E)$ in each energy bin. For the FRONT/BACK division, there are 437 parameters in total. 

To compute the PSF for a specific source in a given sky direction $(RA, Dec)$ at a given energy $E$, one needs to sum the King functions (\ref{eq:king}) over different off-axis angles, with weights proportional to the expected number of events from the source with different off-axis angles. This number of events as a function of the off-axis angle can be estimated based on the history of telescope pointings (the "spacecraft file" in Fermi/LAT data terminology). Such weighting would produce correct PSF prediction for a source that is not variable in time. Otherwise, a bursting source for which most of the event statistics are accumulated during a short outburst may have a PSF that is very different from the steady-state source PSF, because most of the source signal would be composed of events incident at a specific off-axis angle at which the source was visible during the outburst. This suggests a different approach for determination of the PSF for a bursting source (this approach would also work for a steady-state source). The off-axis angle weighting in the PSF calculation can be done based on the real statistics of events from the source, for example, the events within certain limiting angular distance from the source position.

In general, the uncertainties of the PSF parameters include statistical and systematic uncertainties. The statistical uncertainties are determined by the statistics of the Monte-Carlo event sample and they can in principle be reduced via increase of the Monte-Carlo event sample size. To the contrary, the systematic uncertainties cannot be assessed and cannot be reduced using the Monte-Carlo data alone. There may be certain unaccounted instrumental effects that are not considered in Monte-Carlo simulations, the telescope hardware may have characteristics that evolve with time in a way that is not completely controlled etc. Hence, the systematic uncertainty of the PSF has to be addressed via comparison with real data. Such a comparison should ideally be based on the data on genuine point sources detected by Fermi/LAT. 

\section{PSF study with pulsars}
\label{sec:pulsars}

The point source signals are naturally provided by pulsars, fast rotating neutron stars. In general, pulsar signals are superimposed onto extended emission from its pulsar wind nebula and on diffuse Galactic emission. There may also be weak sources in the direct vicinity of the pulsar and it may be challenging to deconvolve emission from these weak sources from the pulsar emission. In this sense, pulsar total emission does not represent a "clean" point source. However, the extended pulsar wind nebula flux, the weak nearby point sources and the diffuse Galactic emission signals do not vary periodically and are identical during the "on-pulse" and "off-pulse" phases of the pulsed signal. The only difference in the signal between the two phases is the presence of the flux coming from the direct vicinity of the neutron star, within the distance of the "light cylinder" which is clearly not resolvable by Fermi/LAT for any of the known pulsars. Subtracting the "off-pulse" flux from the "on-pulse" flux automatically removes the pulsar wind nebula, the weak nearby sources and the diffuse Galactic emission. It is well possible that the "off-pulse" signal also contains weaker emission from the pulsar magnetosphere and subtraction of the "off-pulse" flux can also remove part of the point source flux. This reduction of the point source flux reduces the statistics of the "pure" point source signal, but does not affect its morphology. Thus the off-pulse subtracted signal from pulsars provides a genuine point source flux that can be used for study of the systematic uncertainty of the PSF.

We use pulsars to study the PSF models provided in Fermi/LAT IRFs. We do this for the specific event class P8R3\_SOURCEVETO\_V3. This event selection is characterized by the lowest residual charged cosmic ray background and is therefore most suitable for the studies of weak extended sources (as specified in recommendations on data selections by Fermi Science Support Center, https://fermi.gsfc.nasa.gov/ssc/data/analysis/scitools/ lat\_data\_selection.html). The characteristics of the Monte-Carlo derived PSF for this event class are given in the psf\_P8R3\_SOURCEVETO\_V3\_PSF.fits (for the PSF0-PSF3 event grouping) and psf\_P8R3\_SOURCEVETO\_V3\_FB.fits (for the FRONT-BACK event grouping) files of Fermi/LAT IRF set for the Pass 8, Version 3 data (P8V3). 

Our analysis considers stacked signal of the three brightest pulsars on the sky, Vela, Geminga and Crab. We select events of different event types, PSF0,...,PSF3, FRONT, BACK, by filtering LAT event files using {\it gtselect}-{\it gtmktime} tools chain (see https://fermi.gsfc.nasa.gov/ssc/data/analysis/) specifying the event types that correspond to different event selections. We retain only events in the energy range above 100~MeV. We limit our data-PSF model comparison to the regions within $5^\circ$ around the pulsar position. We consider events up to maximum Zenith angle $100^\circ$. At the {\it gtmktime} step, we impose a filter (DATA\_QUAL\>0)\&\&(LAT\_CONFIG==1), as recommended by the Fermi Science Support Center (https://fermi.gsfc.nasa.gov/ssc). We also limit the time span of the data for each pulsar according to the time limits of validity of available pulsar ephemerides \citep{Fermi-LAT:2023zzt}.

\begin{figure}[h!]
    \includegraphics[width=\columnwidth]{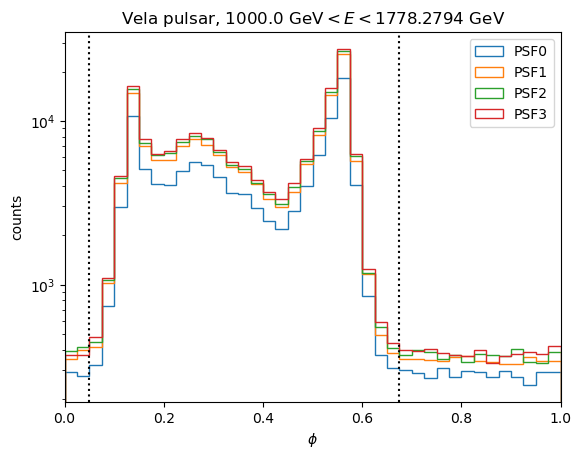}
    \includegraphics[width=\columnwidth]{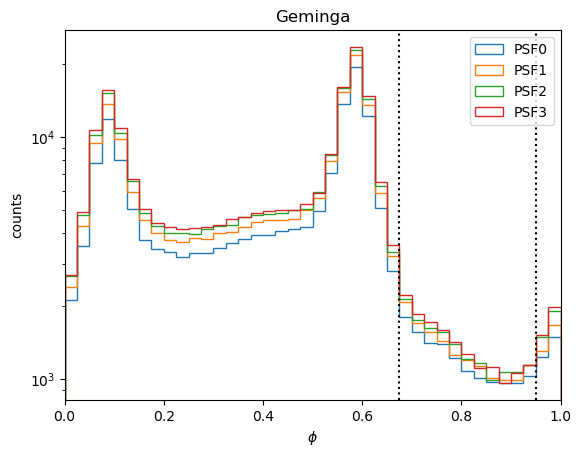}
    \includegraphics[width=\columnwidth]{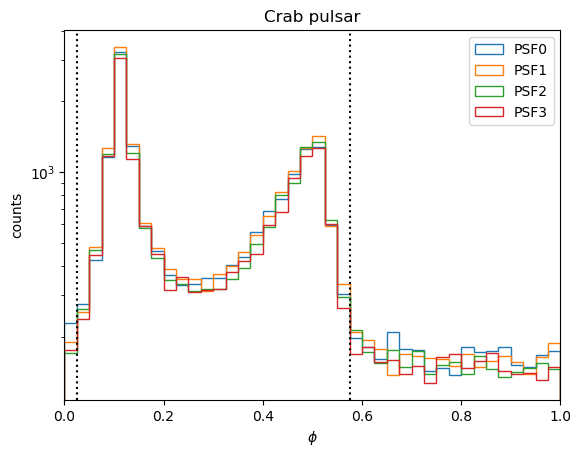}
    \caption{Pulse profiles of Vela, Geminga and Crab pulsars in the energy range 1-1.7 GeV for different event types (PSF0-PSF3). Vertical lines show the boundaries of the on-pulse / off-pulse intervals.}
    \label{fig:pulse_profile}
\end{figure}

To select the "on-pulse" and "off-pulse" signals, we use pulsar ephemerides provided with the 3-rd pulsar catalog of Fermi/LAT \citep{Fermi-LAT:2023zzt}. We use the {\it pint} program \citep{2021PINTprogram} to assign the pulse phase for each event. The ephemerides files are taken from the data repository associated with the pulsar catalog paper (https://fermi.gsfc.nasa.gov/ssc/data/access/lat/ 3rd\_PSR\_catalog/). The current version of {\it pint} is not able to handle the ephemerides file for the Crab pulsar from this repository and we use the {\it tcb2tdb} script to convert the time format of the ephemerides file from "TCB" to "TDB". Plotting the event data as a function of the pulse phase we obtain the pulse profiles as shown in Fig. \ref{fig:pulse_profile} for the energy bin 1-1.7 GeV. Based on these pulse profiles, we define the "on-pulse" and "off-pulse" phases. For Vela, we use the on-pulse phase interval $0.05<\phi<0.675$, for Geminga $-0.05<\phi<0.675$ and for Crab $0.025<\phi<0.575$. The phase interval complementary to the on-pulse is the off-pulse interval used for estimation of the background on top of which the point source signal is observed. 

\begin{figure}
    \includegraphics[width=\columnwidth]{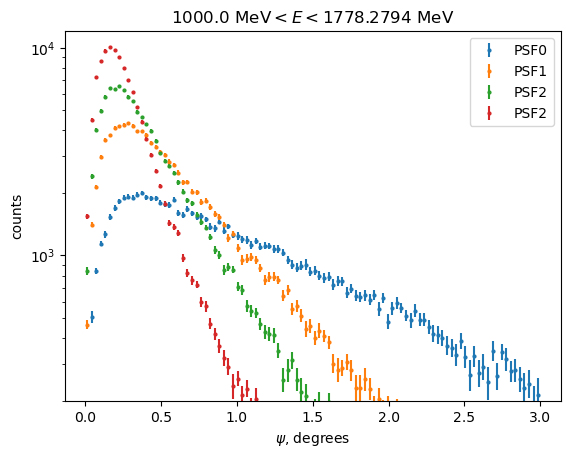}
    \includegraphics[width=\columnwidth]{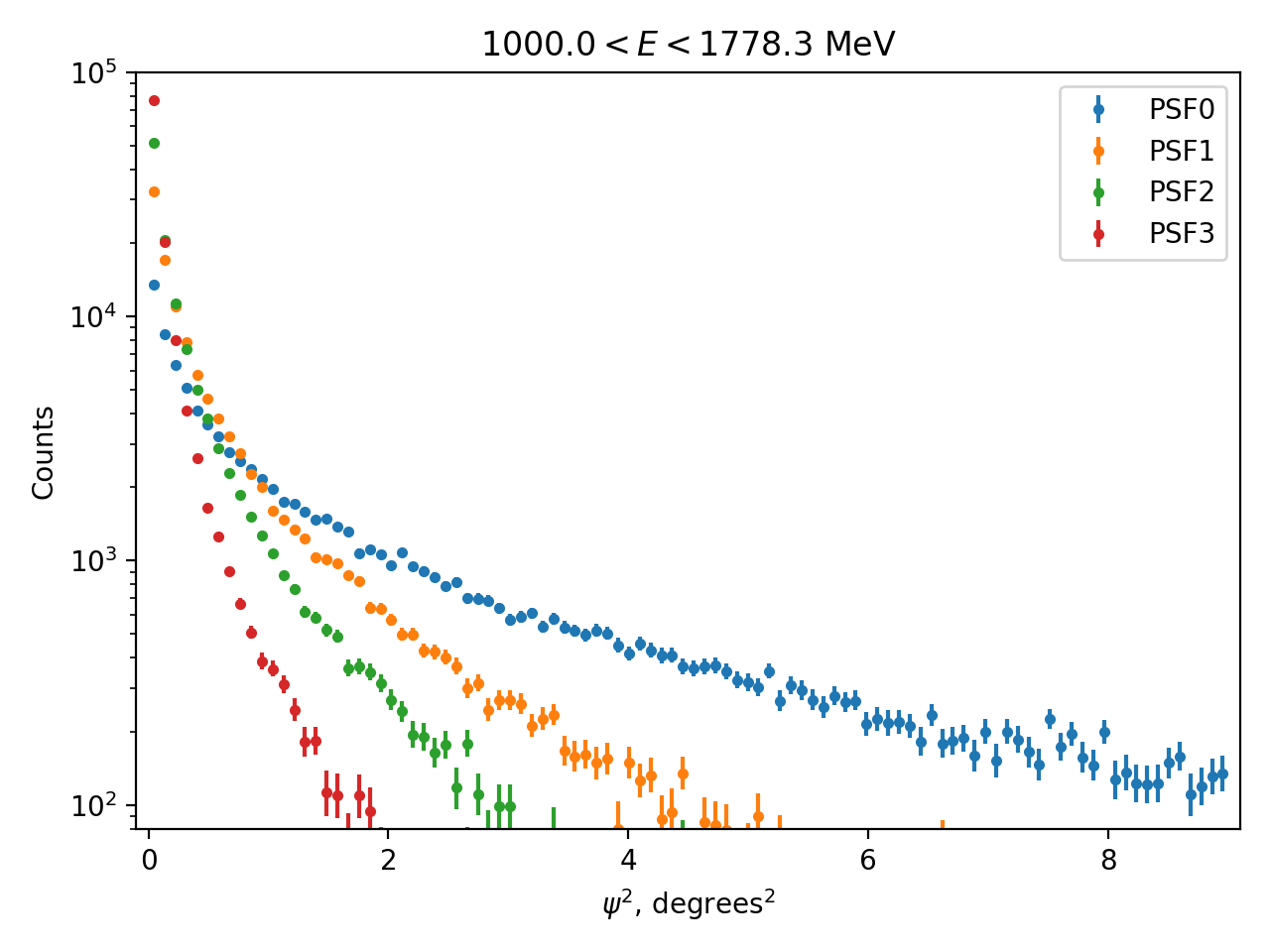}
    \caption{Radial profiles of the stacked pulsed emission signal for different event types (PSF0-PSF3) in the energy range 1-1.7 GeV. Top: binning in equal $\psi$ bins; bottom binning in equal $\psi^2$ (solid angle) bins.}
    \label{fig:psf_types}
\end{figure}

As an example, Fig. \ref{fig:psf_types} illustrates the difference in the event types PSF0,...,PSF3 for different quality of the angular reconstruction. The two panels show events binned in equal angular bins $\psi$ and equal solid angle bins $\psi^2$ respectively. The top panel gives a better insight on the behavior of the inner PSF while the bottom panel provides a view of the "surface brightness" profile of the PSF. One can see that the events belonging to the PSF3 type have the best angular reconstruction quality and are distributed closest to the source. The events from the type PSF0 have the worst PSF quality. The angular width of the PSF for all the event types increases with decreasing energy and below 100~MeV it becomes larger or comparable to the angular cut of $5^\circ$ considered in our analysis. We limit the analysis to $E>100$~MeV for this reason (only bins starting from energy bin 4 of the IRF file are considered). The pulsed emission signal statistics becomes weak in the energy range above 10~GeV and our analysis is effectively limited to the energy range below this energy.

\begin{figure*}
    \includegraphics[width=\columnwidth]{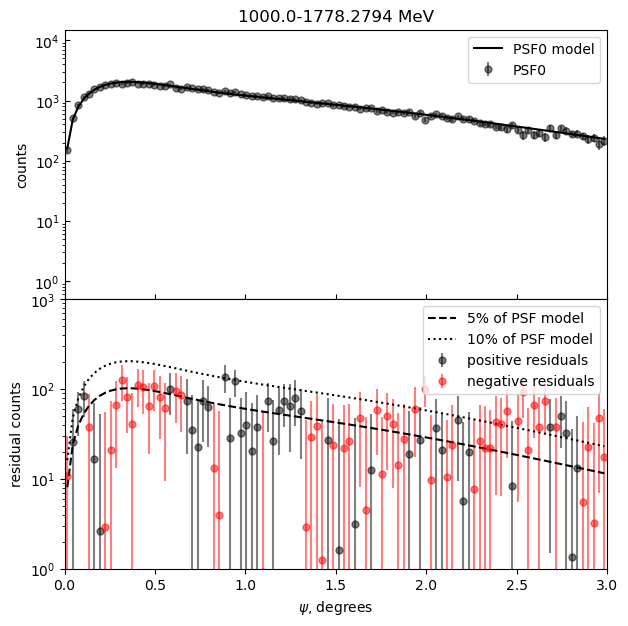}
\includegraphics[width=\columnwidth]{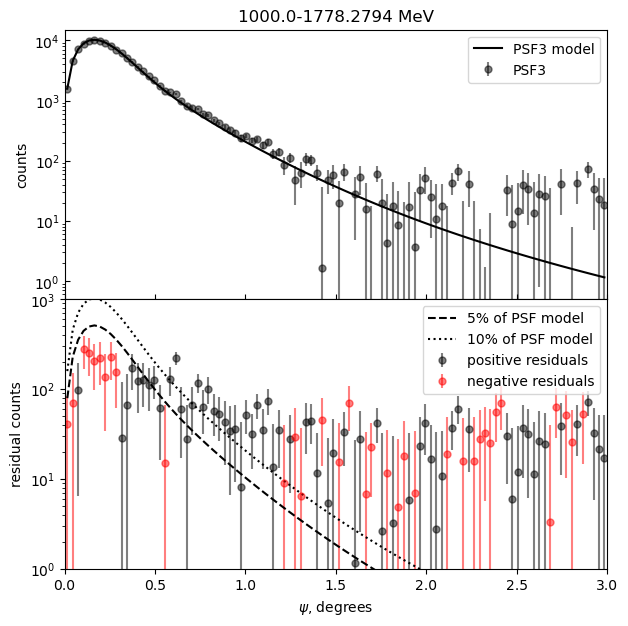}
    \caption{Radial profiles of Vela pulsar signal in the energy range 1-1.7~GeV, for different event types (PSF0, PSF3). Top panels show the fits of the data with the PSF model described in section \ref{sec:psf} (black solid lines). Bottom panels show the positive (black) and negative (red) fit residuals with the residual flux levels corresponding to 5\% (black dashed lines) and 10\% (black dotted lines) of the point source signal.}
    \label{fig:profiles_Vela}
\end{figure*}

\begin{figure*}
    \includegraphics[width=\columnwidth]{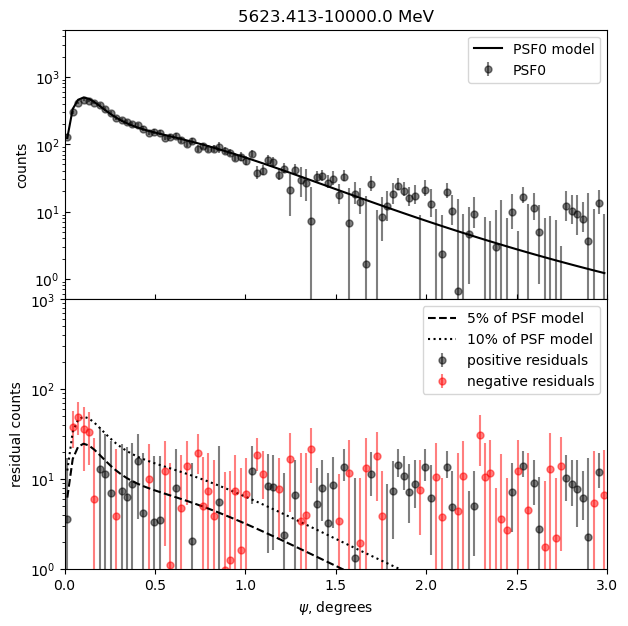}
\includegraphics[width=\columnwidth]{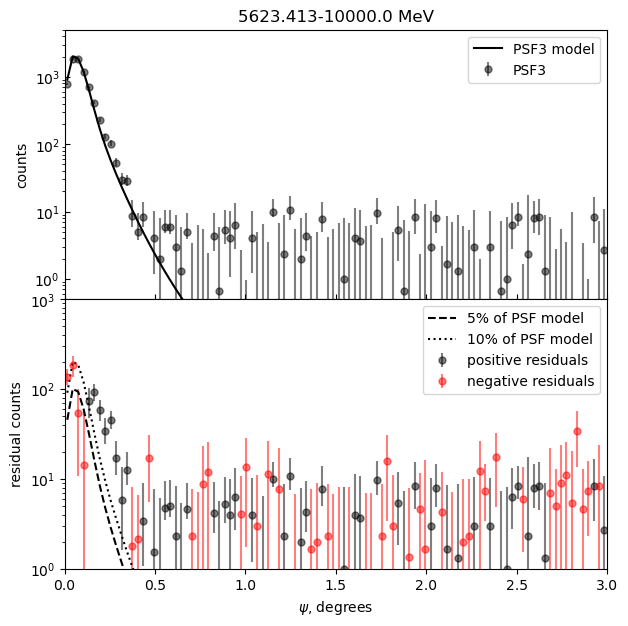}
    \caption{Same as in Fig. \ref{fig:profiles_Vela}, but for the energy range 5.6-10 GeV.}
    \label{fig:profiles_Vela_3GeV}
\end{figure*}

Fig. \ref{fig:profiles_Vela} and \ref{fig:profiles_Vela_3GeV} show comparisons of the model PSF radial profiles from Fermi/LAT IRFs with the pulsar data in the energy ranges 1-1.7 GeV and 5.6-10 GeV for different event types. From the top panels of the figures one can see that the analytical model described in section \ref{sec:psf} provides an overall satisfactory description of the data, in the first approximation. The bottom panels show residuals of the data fits with the PSF models derived from the IRF file, with positive residuals shown in black and negative residuals in red. 

These panels show discrepancies of models with the data, that may reach way beyond 10\% of the PSF over a wide range of angles, as for example in the last panel of Fig. \ref{fig:profiles_Vela_3GeV}. Even though the fraction of the point source signal contained in the PSF tails is small, the data -- model mismatch cannot be ignored in the studies of extended emission, because the mismatch mimics an excess over the point source signals that can be produced by the extended emission.

Deviations of the shapes of the model PSF from the real data can be spotted in the analysis of residuals of the model fits to the pulsar data. Fitting the PSF models to the data for each energy bin, each off-axis angle and each event type, we find that most of the time, the data are not consistent with the model of the PSF given by the IRF and better model fits can be found by fitting the PSF model parameters to the pulsar data. 

\begin{figure*}
    \includegraphics[width=\columnwidth]{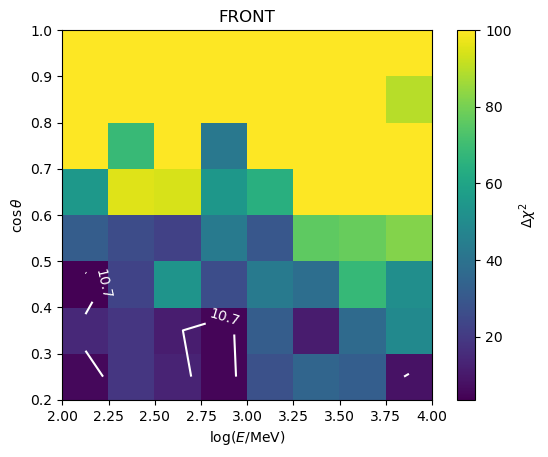}
    \includegraphics[width=\columnwidth]{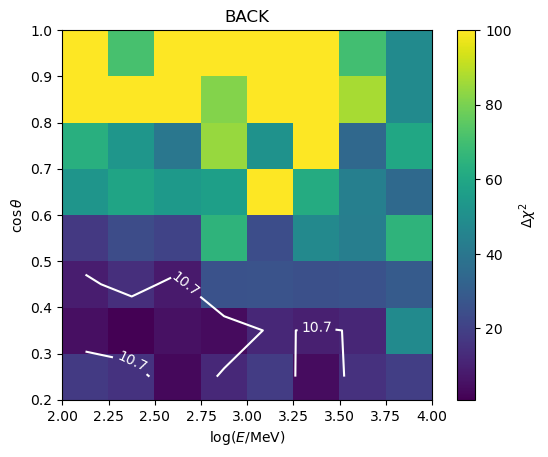}
    \caption{Quality of the fit of the pulsar data with PSF derived from the P8R3 IRFs. Left panel is for the FRONT events and right panel is for the BACK events. Color shows the excess $\chi^2$ of the data fit with the model PSF with parameters from the IRF file, compared to the best-fit model with the Kings function parameters fitted to the data. }
    \label{fig:chi2}
\end{figure*}

Fig. \ref{fig:chi2} shows the difference in the $\chi^2$ between the data fits with the PSF models with parameters extracted from the IRF file and with the five PSF parameters $f_c,\sigma_c,\gamma_c,\sigma_t,\gamma_t$ left free in the fitting. With improvements of $\chi^2$ by as much as $\Delta\chi^2\ge 100$ in more than half of the energy and off-axis angle bins, the PSF parameters given in the IRF file lie outside the $90\%$ confidence range of the $f_c,\sigma_c,\gamma_c,\sigma_t,\gamma_t$ measurements from the pulsar data. The white contours in Fig. \ref{fig:chi2} show these 90\% contours that correspond to $\Delta\chi^2=10.7$ given that the fit adjusts six parameters: the five parameters of the PSF plus the overall normalization of the model \citep{1976ApJ...208..177L}. The mismatch of the PSF model parameters and the data in individual off-axis angle bins is illustrated in Fig. \ref{fig:profile_3_5_7}. 

\section{Improved PSF model based on pulsar data}
\label{sec:psf_revision}

\begin{figure}
    \includegraphics[width=\columnwidth]{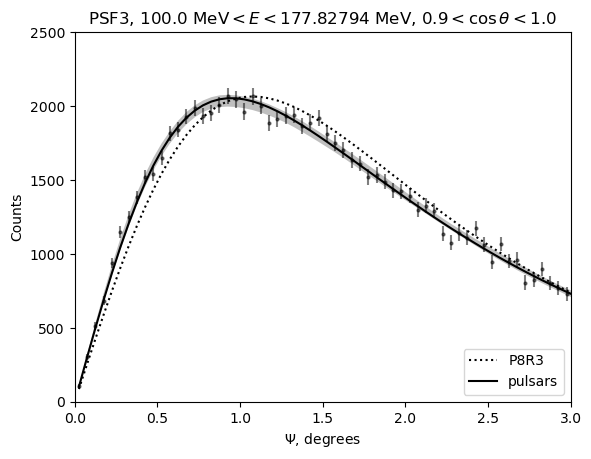}
    \caption{Radial profile of pulsed signal counts for the energy bin 100-177 MeV and off-axis angle bin $0.9<\cos\theta<1$ for the event type PSF3. Dotted line shows the model PSF from Fermi/LAT IRF. Solid line with a gray band shows the best-fit model with $1\sigma$ uncertainty range.}
    \label{fig:profile_3_5_7}
\end{figure}

\begin{figure}
    \includegraphics[width=\columnwidth]{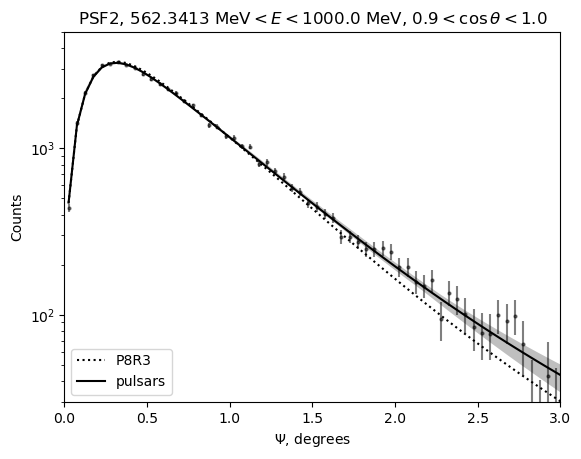}
    \caption{Same as in Fig. \ref{fig:profile_3_5_7} but for the energy range 0.56-1 GeV and off-axis angle bin $0.9<\cos\theta<1$ for the event type PSF2.}
    \label{fig:profile_2_8_7}
\end{figure}

\begin{figure}
    \includegraphics[width=\columnwidth]{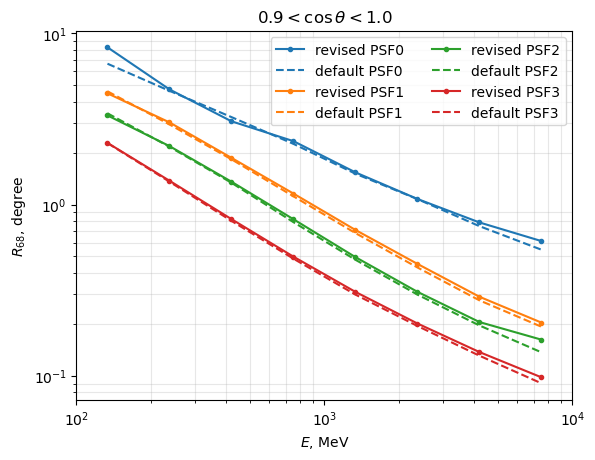}
    \caption{The 68\% containment radius for the event types PSF0 (blue), PSF1 (orange), PSF2 (green) and PSF3 (red) in the energy range 0.1-10 GeV and off-axis angle bin $0.9<\cos\theta<1$. The solid lines and the dashed lines are derived from the revised and default PSF parameters respectively.}
    \label{fig:containment_radius_68}
\end{figure}

Limitations of the PSF models derived from Monte-Carlo studies, described in the previous section, motivate an alternative data-driven approach for modeling of the PSF in the energy range 0.1-10 GeV where high statistics signal from pulsars is available.

One possibility is to use directly the pulsed emission signal, rather than model PSF, as a template for the point source signals. We bin the real pulsar events into the same energy and off-axis angle bins as the IRF file describing the PSF. Data points in 
Figs. \ref{fig:profile_3_5_7}, \ref{fig:profile_2_8_7} show examples of the binned pulsar signals for specific energy, off-axis angle and PSF type bins. One can see that the statistics of the pulsar signal at 100~MeV -- 1 GeV energy range is so high that the error bars are barely noticeable in the profile within $1^\circ-2^\circ$ distance from the source, seen if the data are split into narrow off-axis angle bins. Replacing the analytical models of the PSF with the real-data PSF has an advantage that it completely removes the systematic error of the PSF modeling, including possible error introduced through the choice of the analytical template of the PSF: King function vs. Gaussian, one King function vs. two- or three- King function etc. models.

Using the binned pulsed emission data as PSF templates, it is possible to construct the expected PSF for a source in a specific sky direction, based on the observed distribution of events within certain distance from the source as a function of the off-axis angle, $W(\cos\theta)$. The PSF in a given energy bin is found by summing the lookup table $T$ over the off-axis angle
\begin{equation}
\label{eq:psf_convolution}
PSF(E,\psi)=\sum_{\cos\theta}W(\cos\theta)T_{pulsar}(E,\cos\theta,\psi)
\end{equation}

\begin{table*}
\centering
\begin{tabular}{|l||l|l|l|l|l|l||l|l|l|l|l|l|l|l|l|}
\hline
\multicolumn{13}{| c |}{$1$~GeV$<E<1.7$~GeV}\\
\hline
 $\cos\theta$&$f_{c}$ & $\sigma_c$ & $\gamma_c$ & $\sigma_t$ &$\gamma_t$&$\chi^2$&$f_c$&$\sigma_c$ & $\gamma_c$ & $\sigma_t$ &$\gamma_t$&$\chi^2$\\
 \hline
 \hline
&\multicolumn{12}{| c |}{FRONT}\\
\hline
&\multicolumn{6}{| c ||}{P8\_R3 IRF}&\multicolumn{6}{ c |}{pulsars} \\
\hline
0.2 - 0.3 & 0.72 & 0.55 & 1.69 & 1.93 & 4.87 & 140.36 & 0.72 & 0.54 & 1.53 & 11.14 & 4.59 & 113.02 \\
\hline
0.3 - 0.4 & 0.71 & 0.42 & 2.18 & 1.34 & 2.60 & 124.22 & 0.64 & 0.43 & 2.14 & 1.43 & 2.18 & 92.22 \\
\hline
0.4 - 0.5 & 0.79 & 0.36 & 2.26 & 1.26 & 2.68 & 154.44 & 0.74 & 0.33 & 1.96 & 1.01 & 2.20 & 111.04 \\
\hline
0.5 - 0.6 & 0.78 & 0.34 & 2.46 & 1.16 & 2.77 & 145.13 & 0.76 & 0.34 & 2.44 & 1.17 & 2.71 & 115.36 \\
\hline
0.6 - 0.7 & 0.76 & 0.33 & 2.96 & 1.02 & 2.63 & 152.18 & 0.72 & 0.33 & 2.90 & 1.00 & 2.55 & 87.92 \\
\hline
0.7 - 0.8 & 0.77 & 0.33 & 3.18 & 1.01 & 2.72 & 219.57 & 0.74 & 0.34 & 3.18 & 1.00 & 2.83 & 86.42 \\
\hline
0.8 - 0.9 & 0.74 & 0.33 & 3.92 & 0.91 & 2.64 & 231.34 & 0.71 & 0.33 & 3.73 & 0.89 & 2.61 & 115.57 \\
\hline
0.9 - 1.0 & 0.78 & 0.32 & 3.43 & 0.93 & 2.47 & 248.46 & 0.77 & 0.33 & 3.36 & 0.92 & 2.55 & 126.52 \\
\hline
\hline
&\multicolumn{12}{| c |}{BACK}\\
\hline
&\multicolumn{6}{| c ||}{P8\_R3 IRF}&\multicolumn{6}{ c |}{pulsars} \\
\hline
0.2 - 0.3 & 0.44 & 0.54 & 2.69 & 1.23 & 3.18 & 143.99 & 0.44 & 0.55 & 2.48 & 0.85 & 1.54 & 125.40 \\
\hline
0.3 - 0.4 & 0.56 & 0.43 & 2.43 & 1.14 & 3.26 & 110.30 & 0.56 & 0.39 & 2.09 & 1.08 & 2.70 & 98.30 \\
\hline
0.4 - 0.5 & 0.62 & 0.35 & 2.45 & 1.08 & 3.46 & 105.68 & 0.62 & 0.30 & 1.78 & 0.78 & 2.51 & 79.64 \\
\hline
0.5 - 0.6 & 0.65 & 0.29 & 2.33 & 0.96 & 3.32 & 106.63 & 0.64 & 0.28 & 2.09 & 0.87 & 2.86 & 82.27 \\
\hline
0.6 - 0.7 & 0.70 & 0.27 & 2.42 & 0.93 & 3.36 & 199.72 & 0.66 & 0.25 & 1.97 & 0.76 & 2.67 & 97.94 \\
\hline
0.7 - 0.8 & 0.74 & 0.25 & 2.32 & 0.86 & 3.14 & 187.71 & 0.70 & 0.25 & 2.26 & 0.84 & 3.08 & 136.38 \\
\hline
0.8 - 0.9 & 0.83 & 0.26 & 2.35 & 1.10 & 4.00 & 285.29 & 0.82 & 0.25 & 2.06 & 0.88 & 3.81 & 105.03 \\
\hline
0.9 - 1.0 & 0.81 & 0.25 & 2.53 & 0.93 & 3.20 & 333.15 & 0.78 & 0.25 & 2.29 & 0.81 & 2.96 & 99.03 \\
\hline
\end{tabular}
\caption{Example of revision of PSF parameters for the energy bin $1$~GeV$<E<1.7$~GeV. The two groups of columns for each $\cos\theta$ bin show the parameters of the PSF described in Section \ref{sec:psf} from the IRF file (columns marked "P8\_R3") and the parameters found by fitting the pulsar data (columns marked "pulsars"). The last column of each group shows the $\chi^2$ of the fit of the data with the model PSF.}
\label{tab:psf_lookup}
\end{table*}

An alternative possibility is to re-use and improve the analytical model of the PSF described in section \ref{sec:psf} by fitting the parameters $f_c,\sigma_c,\gamma_c,\sigma_t,\gamma_t$ to the pulsed emission data. An example of such fitting is shown in Fig. \ref{fig:profile_3_5_7}, \ref{fig:profile_2_8_7} for a specific energy and off-axis angle bins by solid lines surrounded by gray shadings that show the 68\% uncertainty ranges of the model fits. To calculate these error bands, we have run Monte-Carlo simulations randomly displacing the PSF parameters from the best fit and noticing the PSF models for which the $\chi^2$ of the model fit does not exceed the best-fit model $\chi^2$ by more than $\Delta\chi^2=7.04$ (which corresponds to the $68\%$ confidence range for the six fitted parameters \citep{1976ApJ...208..177L}). These improved PSF models are compared with the initial PSF models from the IRF. Different types of the model-data discrepancies can be spotted for the PSF models from the IRF: a mismatch in the core of the PSF is visible in Fig. \ref{fig:profile_3_5_7}, a mismatch in the PSF tail is seen in Fig. \ref{fig:profile_2_8_7}. There are no such discrepancies in the re-fitted models that match the data by definition. 

As mentioned in the Introduction, PSF models can differ in shape (as shown in Fig. \ref{fig:profile_2_8_7} and Fig. \ref{fig:profile_3_5_7}) but can have the same  68\% containment radii, meaning that the containment radius is possibly not the only parameter useful for the model validation. This is illustrated by Fig. \ref{fig:containment_radius_68} from which one can see that the 68\% containment radii for both PSF agree with each other up to few percent.

Table \ref{tab:psf_lookup} shows an example comparison of the original and revised PSF parameter tables for a specific energy bin, $1-1.7$~GeV, for the FRONT/BACK event grouping. The table also gives the $\chi^2$ of the pulsar data fits with the model PSF from the IRF file and best-fit model.

An advantage of the analytical PSF model compared to the stacked data PSF model is that it provides a smaller error at large angular distances from the source. This is clearly seen from Figs. \ref{fig:profile_3_5_7}, \ref{fig:profile_2_8_7}: the gray uncertainty bands of the analytical models are narrower than the scatter of the stacked signal data points at large angular distances from the source. A disadvantage of the analytical PSF models is the presence of a residual systematic error due to the fact that the analytical model does not necessarily provide the perfect description of the PSF shape. For the rest of our study, we consider the analytical PSF model only.

\section{Application of revised PSF to the search of extended emission around Mrk 501}
\label{sec:mrk}

Mrk 501 is the second-brightest blazar on the TeV sky \citep{LHAASO:2023rpg}. Detectability of extended secondary emission around this source has been studied in \cite{Korochkin:2020pvg}, which concluded that magnetic field as strong as $10^{-11}$~G is measurable through the search of the extended secondary emission in the energy range up to $E>100$~GeV. The range of weaker magnetic field strengths can be explored via the search of the secondary signal at lower energies accessible to Fermi/LAT, the idea that has been exploited in \cite{Webar:2025qbp}. It is worth noticing that Mrk 501 is not the best source for the search of the extended emission in the energy range of Fermi/LAT, because it has a very strong point source signal in 1-100~GeV range. Sources with harder spectra, like the extreme blazars 1ES 0229+200 or 1ES 0502+675 provide better sensitivity because of lower ratio between the expected point source and extended emission fluxes \citep{Blunier:2025ddu}.

The secondary flux at the energy $E_\gamma\sim 10$~GeV is produced following interactions of $\gamma$-rays with energies
\begin{equation}
    E_{\gamma_0}\simeq 3\left[\frac{E_\gamma}{10\mbox{ GeV}}\right]^{1/2}\mbox{ TeV}
\end{equation}
in the intergalactic medium \citep{2009PhRvD..80l3012N}. The mean free path of such $\gamma$-rays is 
\begin{equation}
    D_{\gamma_0}\sim 2\times 10^2\left[\frac{E_{\gamma_0}}{3\mbox{ TeV}}\right]^{-0.5}\mbox{ Mpc}
\end{equation}
in the range $1\mbox{ TeV}<E_{\gamma_0}<10\mbox{ TeV}$ \citep{Franceschini:2017iwq,Saldana-Lopez:2020qzx}. This is larger than the distance to the source at the redshift $z\simeq 0.034$. The secondary emission detectable in Fermi/LAT energy range is accumulated all over the Line-of-Sight (LoS) toward Mrk 501. 

If the IGMF has parameters $B\sim 10^{-15}$~G and $\lambda_B\sim 10$~kpc, as derived in \cite{Webar:2025qbp}, the angular size of the extended source detectable with Fermi/LAT in 1-10~GeV range is expected to scale with energy as \citep{2009PhRvD..80l3012N}
\begin{equation}
\Psi_{ext}=1.2^\circ\left[\frac{B}{10^{-15}\mbox{ G}}\right]\left[\frac{\lambda_B}{10\mbox{ kpc}}\right]^{1/2}\left[\frac{E_\gamma}{10\mbox{ GeV}}\right]^{-3/4}
\label{eq:theta_ext}
\end{equation}
which is comparable to the extent of the Fermi/LAT PSF that has the 95\% containment radius $\Psi_{PSF}\simeq 1^\circ$ at 10~GeV (see {https://www.slac.stanford.edu/exp/glast/groups/canda/ lat\_Performance.htm}) for the event selection used in \cite{Webar:2025qbp}. 

We start our analysis of Mrk 501 by reproducing the results of \cite{Webar:2025qbp} with the original PSF parameterisations with parameters extracted from the IRF files, using \texttt{fermipy} (v1.4.0)\footnote{https://fermipy.readthedocs.io/en/latest/} \citep{Wood:2017fermipy} and \texttt{fermitools} (v2.2.0)\footnote{https://fermi.gsfc.nasa.gov/ssc/data/analysis/} packages. We subsequently replace the PSF from the original IRF files with the updated PSF based on our pulsar analysis and redo the same analysis using the revised PSF. For the energies outside the pulsar data range, we consider the original PSF parameters.

We select Fermi-LAT data from 17-years of operation between March 2008 and January 2026 in the energy range 0.1-300 GeV. We define a $10^{\circ}\times 10^{\circ}$ ROI centered on Mrk 501 with spatial bin size of $0.02^{\circ}$ and 4 bins per energy decade. We follow the recommended data selection of the Fermi-LAT website for extragalactic diffuse analysis of Pass 8 Data (P8R3)\footnote{https://fermi.gsfc.nasa.gov/ssc/data/analysis/scitools/\ lat\_data\_selection.html} and use the P8R3\_SOURCEVETO\_V3 IRF ($\texttt{evclass=2048}$) with FRONT+BACK photons ($\texttt{evtype=3}$) filtered with the corresponding time selection filtering (DATA\_QUAL$>$0)\&\&(LAT\_CONFIG==1). We run the \textit{gtselect-gtbin-gtmktime-gtexpcube2-gtscrmaps} tool chain to initialize the analysis setup. Finally, we use the best-fit synchrotron-self Compton emission model of \cite{Webar:2025qbp} for the spectrum of Mrk 501 defined as
\begin{equation}
\label{eq:Mrk501_int_spec}
 \frac{dN}{dE} = N_0 \left(\frac{E}{E_0}\right)^{-\Gamma}\exp{\left(-\frac{E}{E_c} \right)^{\beta}},
\end{equation}
with $N_0=1.04 \times 10^{-5}~\mathrm{TeV^{-1}s^{-1}cm^{-2}}$, $\Gamma=1.72$, $E_c=0.33~\mathrm{TeV}$, and $\beta=0.37$. 

We perform a likelihood ratio test where we compute the test statistic
\begin{equation}
\label{eq:TS}
 TS=-2(\ln L_0-\ln L_1)~,
\end{equation}
where $\ln L_0$ and $\ln L_1$ are the log-likelihoods to be maximized in the following two hypotheses. 

\begin{figure}
    \includegraphics[width=\columnwidth]{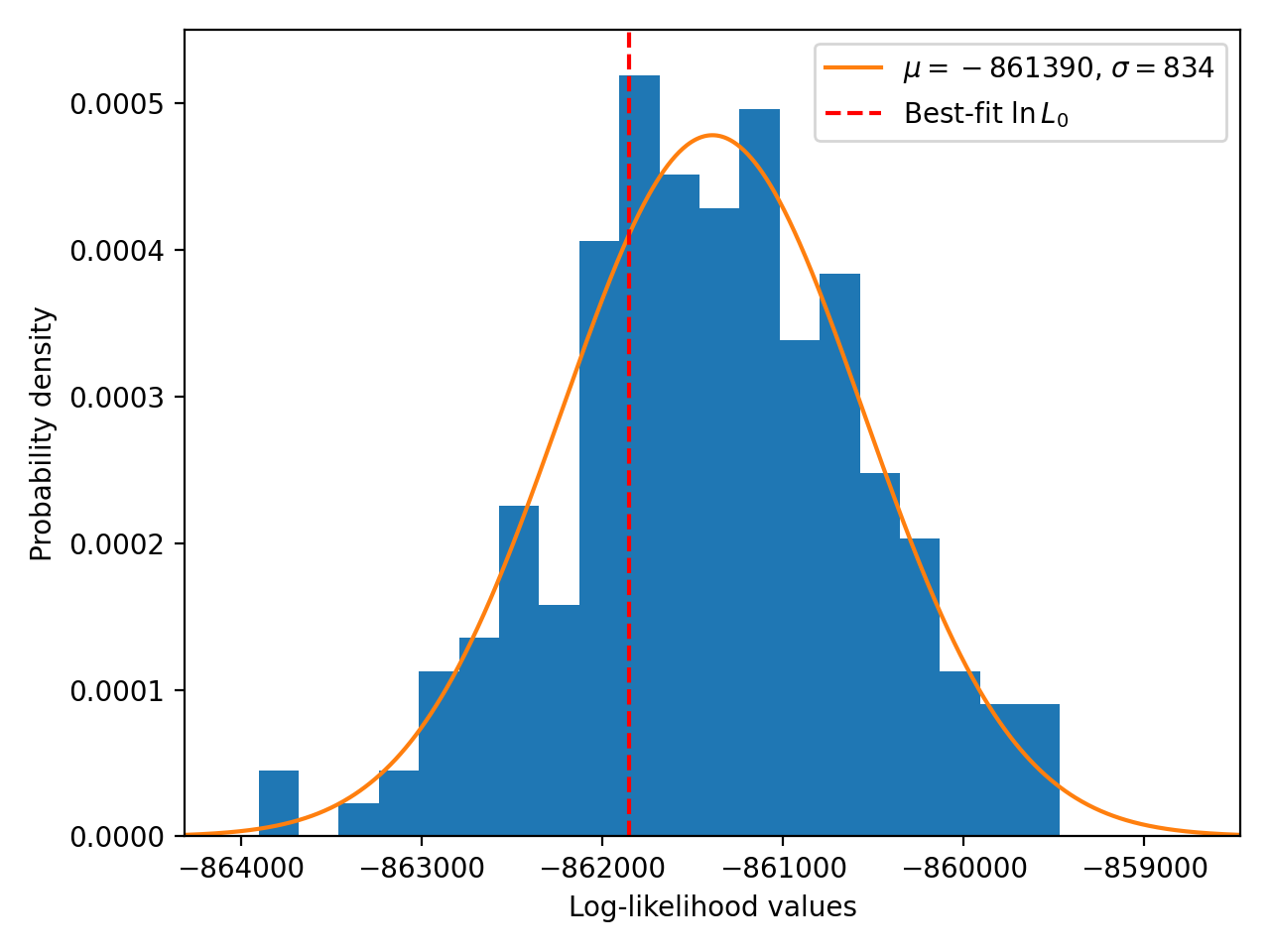}
    \caption{Distribution of the maximal log-likelihood values for the Monte-Carlo simulations of the ROI based on the best-fit model of Mrk 501 ROI. Gaussian distribution with the corresponding mean and standard deviation is presented in orange. Vertical dashed line shows the maximal log-likelihood of the real data fit.}
    \label{fig:likelihood_distribution}
\end{figure}

\begin{figure*}  \includegraphics[width=\columnwidth]{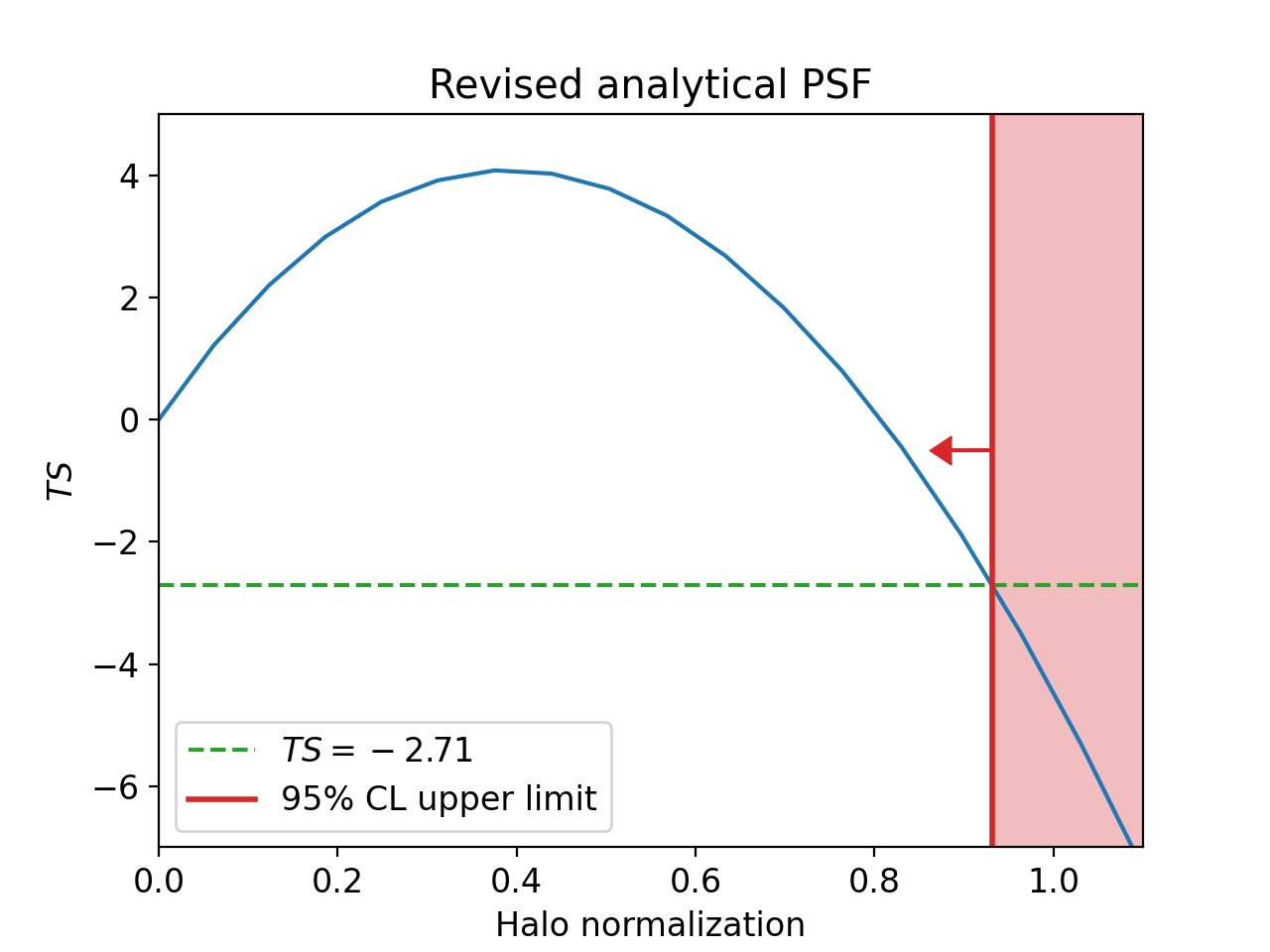}
\includegraphics[width=\columnwidth]{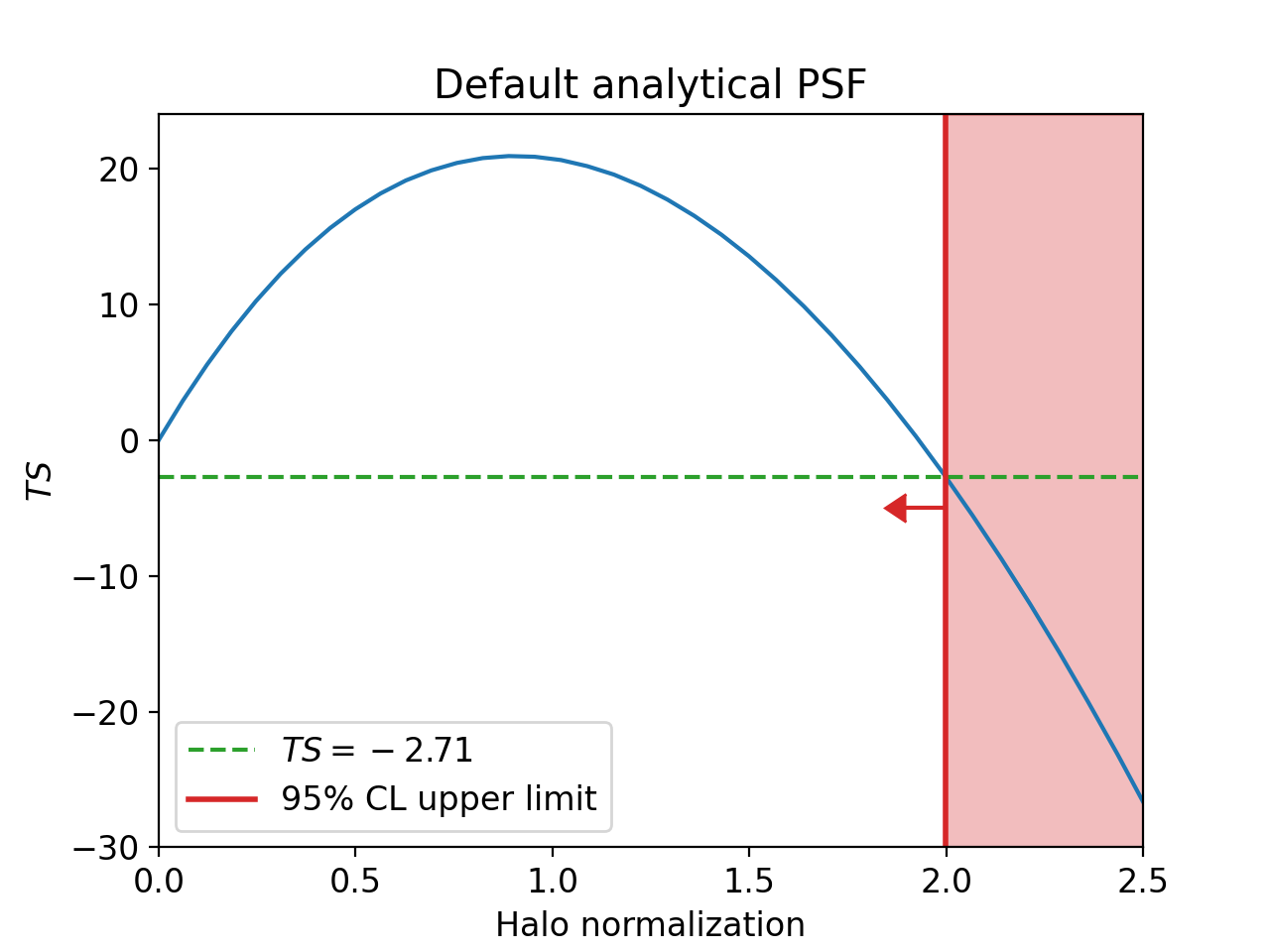}
    \caption{Test statistic profile of halo normalization for the revised (left) and default (right) analytical PSF models. Values above the upper limit on the halo normalization (shaded red) are excluded at the 95\% CL corresponding to $TS=-2.71$ (dashed green).}
    \label{fig:TS_profile_halo}
\end{figure*}

\begin{figure*}   \includegraphics[width=\columnwidth]{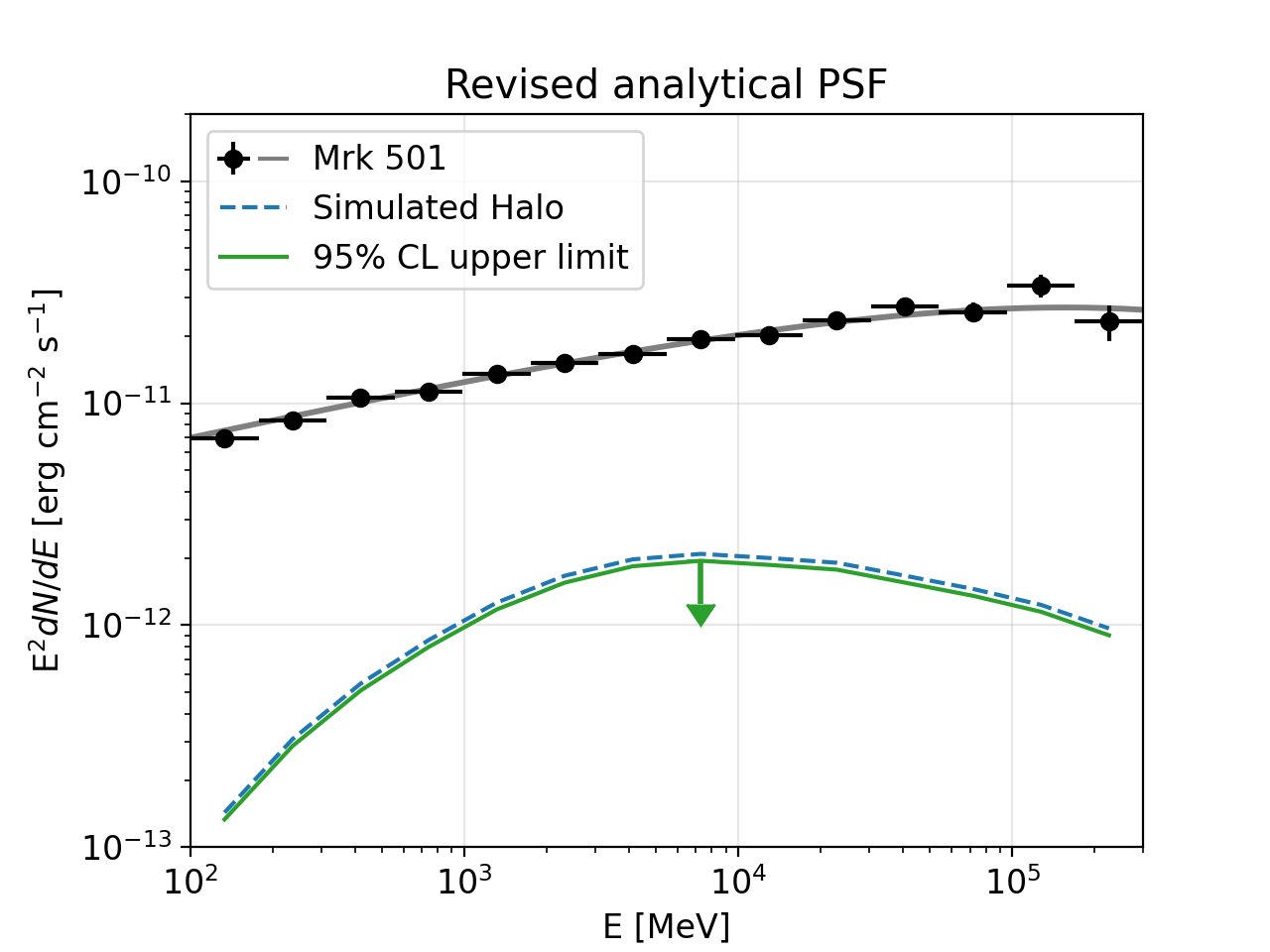}
\includegraphics[width=\columnwidth]{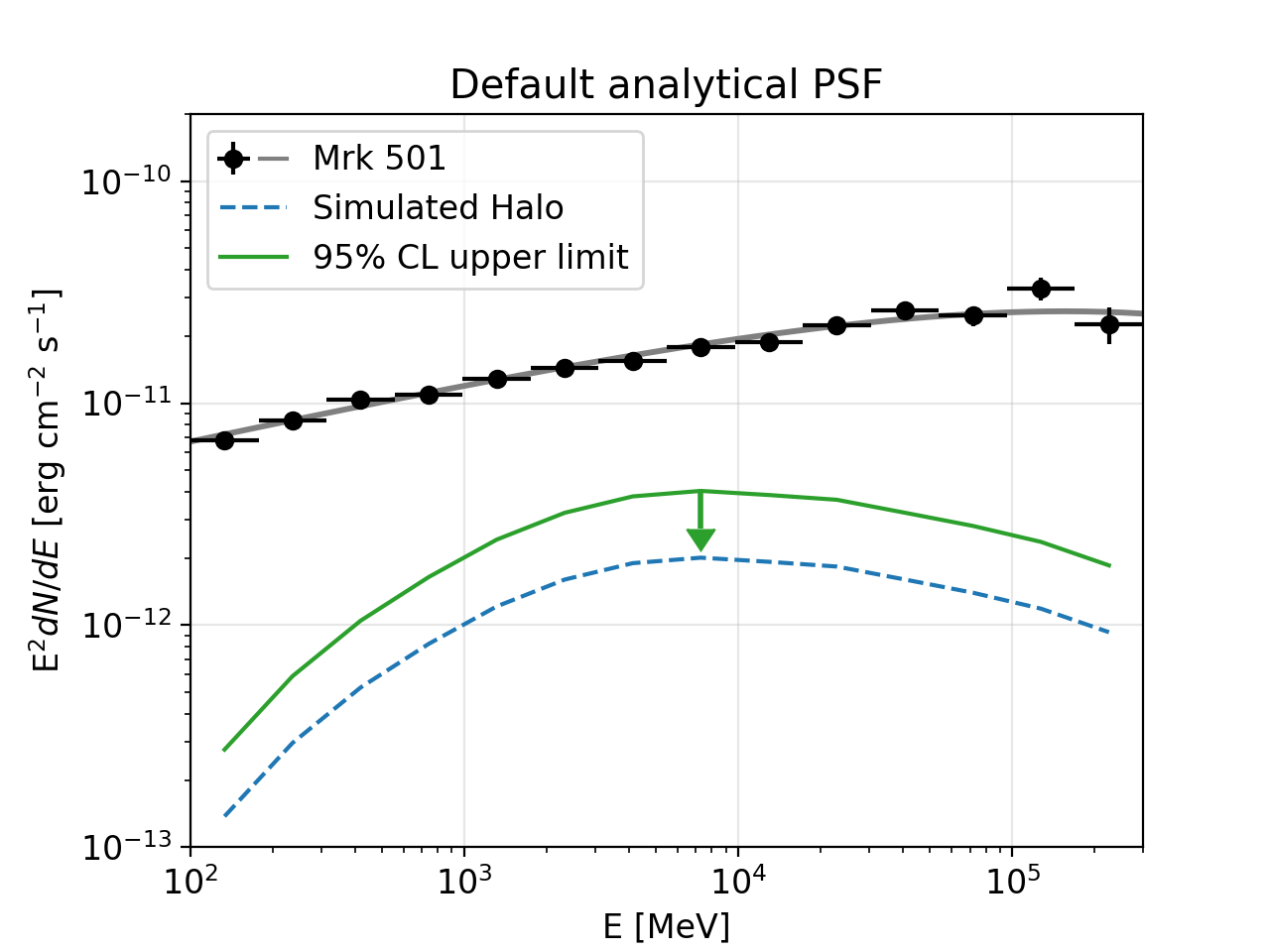}
    \caption{Spectral energy distribution of Mrk 501 (black data points) with best-fit model (gray). The corresponding simulated halo flux from CRbeam (dashed blue) with its 95\% CL upper flux limit (green) are present. Left and right show the results for the revised and default analytical PSFs respectively.}
    \label{fig:sed_mrk501_halo}
\end{figure*}

In the null hypothesis $H_0$, we consider the sky model that contains all sources from the 4FGL-DR4 Fermi catalog \citep{Ballet:20234FGL} within a $15^{\circ}\times 15^{\circ}$ Region-Of-Interest (ROI) centered on Mrk 501. We free the normalization and spectral model parameters of all sources included in a radius of $5^{\circ}$ around Mrk 501 or with a $TS\geq25$. We additionally free the normalization of the isotropic and galactic diffuse models and fix all spectral model parameters of Mrk 501 except its normalization. We fit the ROI model by maximizing the log-likelihood.

In the alternative hypothesis $H_1$, we use the same initial configuration as in $H_0$ and add a source representing the extended emission around Mrk 501 simulated with CRbeam code \citep{Kalashev:2022cja} for a source with the intrinsic spectrum given by Eq. (\ref{eq:Mrk501_int_spec}). The primary gamma-ray source is assumed to emit photons within a jet of $\theta_{\text{jet}}=3^{\circ}$ opening angle. The jet direction is aligned with the line of sight. Secondary photons with a time-delay below $10^7$ years are binned in spatial and energy bins to yield a source template that is used in the likelihood analysis. We allow the normalization of the extended source $\epsilon>0$ to be free in the fit. $\epsilon=1$ corresponds to the secondary flux predicted by the model for specific magnetic field parameters which we take to be the best-fit values of the analysis of \cite{Webar:2025qbp}, $B=1.5\times 10^{-15}$~G and $\lambda_B=10$~kpc. The best-fit $\epsilon$ parameter is rescaled according to the best-fit normalization of Mrk 501.

The analysis based on the default PSF from the IRF files gives $\ln L_0=-861849.18$ and $\ln L_1 = -861838.72$
translating to $TS=20.92$ with $\epsilon=0.89\pm 0.12$, which corresponds to a $\simeq4.57\sigma$ significance of detection of the extended emission. This is somewhat lower than the significance reported in \cite{Webar:2025qbp}. The difference can be due to a difference in the modeling of the extended emission (we use CRbeam code, while the analysis of \cite{Webar:2025qbp} was based on ELMAG code \citep{Blytt:2019ELMAG}) and due to the difference in the event selection: SOURCEVETO photons vs. SOURCE photons.

Once we replace the standard PSF with the PSF derived from the pulsar analysis, we obtain $\ln L_0=-861895.21$ and $\ln L_1 = -861893.17$ corresponding to the best-fit  $\epsilon=0.40\pm 0.12$ with the test statistic $TS=4.08$, so that the extended emission is not significantly detected anymore. To the contrary, the TS value of the point source at the location of  Mrk 501 increases from $TS=49375.54$ to $TS=49579.00$, which hints for a better match between the revised PSF to the point source signal shape.

The maximal likelihood of the fit with the ROI model based on revised PSF is slightly below that of the model based on the IRF PSF. To check if the model based on the revised PSF is consistent with the data, we have performed the "goodness-of-fit" test, by running a set of Monte-Carlo simulations to produce mock Fermi-LAT data realizing the sky model, using \textit{simulate\_roi} function of \texttt{fermipy} package. Fitting the mock data with the reference model, we find a distribution of the log-likelihood values expected in the case when the model is consistent with the data. This distribution is shown in Fig. \ref{fig:likelihood_distribution}. We see that the log-likelihood found for the real data is well within the expected range.

To infer an upper limit on the extended secondary source flux (at 95\% confidence level, CL), we look at a deviation of $\Delta TS=-2.71$ from the $H_0$ hypothesis ($TS=0$) by scanning the $\epsilon$ parameter space and re-fitting the ROI model at each step. The results are shown on the left panel of Fig. \ref{fig:TS_profile_halo} where the maximal halo normalization for the revised analytical PSF model is $\epsilon_{\text{upper}}=0.93$, which is lower than the expected flux normalization $\epsilon=1$. Hence, the extended emission for the best-fit values $B=1.5\times 10^{-15}$, $\lambda_B=10$~kpc of the IGMF is ruled out at the $95\%$ CL. Additionally, we present the results for the default analytical PSF model on the right panel of Fig. \ref{fig:TS_profile_halo} where $\epsilon_{\text{upper}}=2.00$, indicating no disagreement with the presence of the extended emission model from CRbeam code at the 95\% CL. Fig. \ref{fig:sed_mrk501_halo} shows the 95\% CL upper limit on the extended emission flux derived from Fig. \ref{fig:TS_profile_halo} with the default and revised PSF models supporting the same conclusions.

\section{Discussion and conclusions}
\label{sec:discussion}

Knowledge of the telescope PSF is crucial for the search of extended $\gamma$-ray emission around extragalactic sources that may be induced by the effects of propagation of $\gamma$-rays through the intergalactic medium. Use of imprecise PSF model may lead to appearance of false positive detection of extended emission that would be favored by a fit of observational data to "patch" the mismatch between the data and the PSF model in certain angular distance range. Past examples of false positive detections of extended emission around blazars in Fermi/LAT data \citep{Ando:2010rb} have previously led to revision of Fermi/LAT PSF models driven by the model-data comparisons \citep{2011A&A...526A..90N,2013ApJ...765...54A}. 

In this paper we performed the model-data comparison for the PSF models provided by the IRFs of Pass 8, revision 3 event selection of Fermi/LAT data. We have found discrepancies between the models and data on pulsed emission from bright pulsars in the energy range 0.1-10~GeV. The model parameters quoted in the IRF files happen to be inconsistent with the data in different energy and off-axis angle bins. We have revised the PSF modeling for a specific event class, P8R3\_SOURCEVETO, that is most suitable for the search of the extended emission around extragalactic sources.

We have demonstrated the importance of the precise PSF modeling for the search of extended emission around gamma-ray sources by re-assessing the result on detection of extended emission around Mrk 501 \citep{Webar:2025qbp}. Our analysis shows that replacing the PSF models derived from the analytical parameterisations provided by the IRFs with the data-driven PSF "washes out" an evidence for the presence of the extended emission around this source, previously claimed in \cite{Webar:2025qbp}. 

The data-driven approach for the PSF determination has a limitation related to the necessity of high-statistics genuine point source signal provided by pulsars. Such signal is only available in the energy range below $\sim 10$~GeV. Above this energy, the pulsed emission fluxes of pulsars are suppressed and the statistics of the signal drops dramatically. This means that our method for PSF characterization would not work at the highest energies and the Monte-Carlo based PSF determination would still be needed for the extended emission searches up to the highest energies accessible to Fermi/LAT. 

\begin{acknowledgements}
We would like to thank D. Horns for further discussion. This work has been supported in part by the French National Research Agency (ANR) grant ANR-24-CE31-4686. 
\end{acknowledgements}

\bibliographystyle{aa}
\bibliography{refs}

\end{document}